\documentclass[12pt]{article}
\usepackage{a4wide}
\usepackage{amssymb}
\usepackage{amsmath}
\usepackage{bm}

\usepackage{hyperref}
 \numberwithin{equation}{section}
 
\begin{document}

\setcounter{table}{0}

\begin{flushright}\footnotesize

\texttt{ICCUB-18-018}\\
\texttt{Imperial/TP/18/AB/01}
\end{flushright}

\mbox{}
\vspace{0truecm}
\linespread{1.1}


\centerline{\LARGE \bf Universality of Toda equation }
\bigskip

\centerline{\LARGE \bf in ${\cal N}=2$ superconformal field theories}

\vspace{.4cm}

 \centerline{\LARGE \bf }

\vspace{1.3truecm}

\centerline{
    {\large \bf Antoine Bourget${}^{a}$} \footnote{a.bourget@imperial.ac.uk}, 
    {\large \bf Diego Rodriguez-Gomez${}^{b}$} \footnote{d.rodriguez.gomez@uniovi.es}
    {\bf and}
    {\large \bf Jorge G. Russo ${}^{c,d}$} \footnote{jorge.russo@icrea.cat}}

\vspace{1cm}
\centerline{{\it ${}^a$ Theoretical Physics, The Blackett Laboratory}} \centerline{{\it Imperial College London}} \centerline{{\it SW7 2AZ United Kingdom}}
\medskip
\centerline{{\it ${}^b$ Department of Physics, Universidad de Oviedo}} \centerline{{\it Calle Federico García Lorca 18, 33007, Oviedo, Spain}}
\medskip
\centerline{{\it ${}^c$ Instituci\'o Catalana de Recerca i Estudis Avan\c{c}ats (ICREA)}} \centerline{{\it Pg.Lluis Companys, 23, 08010 Barcelona, Spain}}
\medskip
\centerline{{\it ${}^d$ Departament de F\' \i sica Cu\' antica i Astrof\'\i sica and Institut de Ci\`encies del Cosmos}} \centerline{{\it Universitat de Barcelona, Mart\'i Franqu\`es, 1, 08028
Barcelona, Spain }}

\vspace{1.6cm}

\centerline{\bf ABSTRACT}
\medskip

\noindent 

We show that extremal correlators in all Lagrangian ${\cal N}=2$ superconformal field theories with a simple gauge group
are governed by the same universal Toda system of equations,  which dictates the structure of extremal correlators to all orders in the perturbation series.
A key point is the construction of a convenient orthogonal basis for the chiral ring, by arranging towers of operators in order of increasing dimension, which has the property that the associated two-point functions satisfy decoupled Toda chain equations. 
We explicitly verify this in all known SCFTs based on $\mathrm{SU}(N)$ gauge groups as well as in superconformal QCD based on orthogonal and symplectic groups. As a by-product, we find a surprising non-renormalization property for the ${\cal N}=2$ $\mathrm{SU}(N)$ SCFT  with  one hypermultiplet in the rank-2 symmetric representation and one hypermultiplet in the rank-2 antisymmetric representation, where the two-loop terms of a large class of supersymmetric observables identically vanish.

\newpage

\tableofcontents

\section{Introduction and Summary}

Correlation functions are central quantities in Quantum Field Theory. While in general one has to resort to sophisticated perturbative methods to compute them at multiloop level, 
in supersymmetric theories some correlation functions can be computed exactly, thus offering a unique window into the structure of gauge field theories. This is, in particular, the case of extremal correlation functions of chiral primary operators (CPOs) in four-dimensional superconformal field theories (SCFTs) with $\mathcal{N}=2$ supersymmetry and freely generated chiral rings
\cite{Papadodimas:2009eu,Baggio:2014ioa,Baggio:2014sna,Baggio:2015vxa,Gerchkovitz:2016gxx}. 

CPOs are operators annihilated by all the Poincar\' e supercharges of one chirality. It has been argued that CPOs are all scalar operators. Moreover, in SCFTs, because of R-charge conservation, the structure constants of the OPE are true constants and hence the operators are endowed with a ring structure, the so-called chiral ring. Because of this, 2- and 3-point functions contain all the information about the ring. In the following we will be interested on 2-point functions, 
\begin{equation}
\langle O_I(x)\overline{O}_{\overline{J}}(0)\rangle=\frac{g_{I\overline{J}}}{|x|^{\Delta_I+\Delta_{\overline{J}}}}\,\delta_{\Delta_I,\Delta_{\overline{J}}}\, , 
\end{equation}
where  $g_{I\overline{J}}$ is in general a non-trivial function of the Yang-Mills coupling constant.

The chiral ring is an interesting object \textit{per se} with possibly a very rich geometry. Indeed, very recently it has been found that such rings need not  be freely generated \cite{Bourget:2018ond,Argyres:2018wxu,Bourton:2018jwb}.
In this paper, we will consider theories with freely generated chiral rings, where there are no constraints on the CPOs. 

Given an SCFT, the space of all possible marginal deformations defines the conformal manifold. It turns out that   exact marginal deformations preserving $\mathcal{N}=2$ supersymmetry are CPOs of dimension 2. Thus, the matrix $g_{I\overline{J}}$ restricted to the $I,\,\overline{J}$ corresponding to dimension 2 CPOs, known as Zamolodchikov metric, has the interpretation of a metric on the conformal manifold. It is possible to show that the K\" ahler potential for this metric coincides, up to a numerical factor, with the logarithm of the $\mathbb{S}^4$ partition function \cite{Gerchkovitz:2014gta}, which in turn can be exactly computed through localization \cite{Pestun:2007rz}. This implies an important connection between the conformal manifold, the sphere partition function and the correlation functions of CPOs, which was beautifully explored in a series of papers starting with \cite{Papadodimas:2009eu}. For the purpose of this note, it will suffice to recall that extremal correlators of CPOs can be exactly computed through localization as described in \cite{Gerchkovitz:2016gxx}. A crucial insight in  \cite{Gerchkovitz:2016gxx} is that the $\mathbb{R}^4\leftrightarrow \mathbb{S}^4$ mapping induces, through the conformal anomaly, a mixing among operators of any chosen $\Delta$ with lower dimensional ones. This mixing can be disentangled through C (GS) orthogonalization with respect to all lower dimensional operators.

Elaborating on this, there has recently been conceptual and technical progress in understanding different properties of correlation functions of CPOs. 
This includes, in particular, exact results in the large $N$, planar limit \cite{Rodriguez-Gomez:2016ijh,Baggio:2016skg,Rodriguez-Gomez:2016cem,Pini:2017ouj},
where instanton contributions are exponentially suppressed, comparison between localization and field theory computations \cite{Billo:2017glv} and correlation functions involving  Wilson loop operators \cite{Rodriguez-Gomez:2016cem,Billo:2018oog,Giombi:2018qox}.
More recently, extremal correlators have been computed in a novel large $R$ charge, double-scaling limit \cite{Bourget:2018obm}, further investigated in \cite{Beccaria:2018xxl} by exploiting the Toda equation. In some cases, the large $R$ charge limit was shown to admit a precise effective field theory description \cite{Hellerman:2017sur,Hellerman:2018xpi}. 
 
In \cite{Papadodimas:2009eu} it was shown that, when regarding the 2-point correlators as functions of the marginal couplings, a beautiful structure analogous to the 2d $tt^*$ equations emerges. While these equations hold on general grounds, in the following we will concentrate on Lagrangian theories based on simple gauge groups
(with freely generated chiral ring), for which there is only one marginal coupling, namely the Yang-Mills coupling itself $\tau=\frac{\theta}{2\pi}+i\frac{4\pi}{g^2} $. 
In the case of $\mathrm{SU}(2)$ superconformal QCD (SQCD), it was shown in  \cite{Papadodimas:2009eu,Baggio:2015vxa} that the underlying structure is that of a semi-infinite Toda chain. This structure naturally emerges from the exact computation of the correlation functions through localization as a consequence of the algebraic structure of the Gram-Schmidt orthogonalization together with the simple fact that the partition function, discarding instantons, depends on $\tau, \bar\tau$ only through the classical contribution $Z_{cl}=e^{-2\pi\,{\rm Im}\tau\,{\rm Tr}\phi^2}$ (here $\phi$ is the scalar in the vector multiplet), so that derivatives with respect to $\tau, \bar\tau$ give rise to insertions of ${\rm Tr}\phi^2$ (we refer to \cite{Gerchkovitz:2016gxx} for further details). 

The decoupled Toda chain structure was suggested to extend to $\mathrm{SU}(N)$ SQCD based on a  two-loop computation in \cite{Baggio:2015vxa}. In this paper, this was conjectured to hold true to arbitrary loops provided a sufficient condition, which amounts to a certain no-mixing ansatz, is satisfied. This condition was in turn proved not to hold beyond two loops by explicit calculation in the $\mathrm{SU}(3)$, $\mathrm{SU}(4)$ cases in \cite{Gerchkovitz:2016gxx}, concluding that the decoupled Toda chain structure would fail already at three loops. However, very recently, it was shown in \cite{Bourget:2018obm} that, at least for the subsector of operators of the form $({\rm Tr}\phi^2)^n$, the extremal correlators do satisfy a Toda equation. This fact was further explored and extended to the next family of operators of the form $({\rm Tr}\phi^2)^n{\rm Tr}\phi^3$ in \cite{Beccaria:2018xxl}, where it was argued that also for the next one $({\rm Tr}\phi^2)^n{\rm Tr}\phi^4$ a decoupled Toda chain would be obtained upon full orthogonalization. In this paper we show that these partial hints are actually manifestations of the fact that indeed, all extremal correlators of CPOs do satisfy  decoupled Toda chain equations. The key in establishing this fact is that 
there is a natural order for such orthogonalization. To implement it, we consider as seeds the set of all CPOs $O_I$ not including those of the form $({\rm Tr}\phi^2)^n$. The latter play a special role allowing to construct towers starting from the seeds $O_I$ as $T_I=\{O_I,\,O_I\,({\rm Tr}\phi^2),\,O_I\,({\rm Tr}\phi^2)^2,\,\cdots\}$. The prescription is then to orthogonalize each tower with respect to all lower dimensional towers, and, within  any given tower, orthogonalize operators sequentially from left to right. Note that this does not fix the order whenever two towers have the same seed dimension. However it turns out that the order chosen to orthogonalize these does not matter, as any order gives rise to the same decoupled Toda chain. We refer to section \ref{algorithm} below for the detailed prescription. Once this is done, unnormalized extremal correlators of CPOs  satisfy the decoupled Toda chain equations\footnote{We use a multi-index notation explained in section \ref{algorithm}. In particular $\mathbf{1}=(1,0,\dots , 0)$. }
\begin{equation}
\label{TodaGeneral}
\partial_{\tau}\partial_{\bar{\tau}}\log\widetilde{G}_I=\frac{\widetilde{G}_{I+\mathbf{1}}}{\widetilde{G}_{I}}-\frac{\widetilde{G}_{I}}{\widetilde{G}_{I-\mathbf{1}}}\,. 
\end{equation}
These equations in fact encapsulate an infinite number of semi-infinite Toda chains and, consequently,  they need to be supplemented with appropriate boundary conditions. We argue that {\it the same decoupled Toda system \eqref{TodaGeneral} dictates the structure of the perturbation series of extremal correlators in any Lagrangian CFT based on a simple gauge group}. 

We have explicitly checked this fact in all SCFTs based on $\mathrm{SU}(N)$ gauge groups (as classified in \cite{Koh:1983ir}) as well as in SQCD and $\mathcal{N}=4$ SYM based on $SO$ and $Sp$ groups at least to the first term non-linear in Riemann $\zeta$ functions. When computationally possible, we have checked the decoupled Toda chain equations beyond 10 loops. The decoupled Toda chains that we find are precisely those in eq. \eqref{TodaGeneral} in all cases.

A connection between the Toda equation and matrix models is known from very early times \cite{Gerasimov:1990is}. It might   be 
illuminating to adapt this connection to the present case 
of  extremal correlators in $\mathcal{N}=2$ SCFTs. 


Finally, in section \ref{classif} we  point out an intriguing  non-renormalization property for the class of $\mathrm{SU}(N)$
${\cal N}=2$ SCFTs with matter consisting in  one hypermultiplet in the rank-2 symmetric representation and one hypermultiplet in the rank-2 antisymmetric representation: the two-loop terms of 
all extremal correlators
exactly vanish. The corresponding perturbative expansions do  not contain terms with $\zeta(3)$ coefficients.

\medskip
Our main result, namely that for any theory the same system of decoupled Toda chains governs the perturbative expansion of correlators, opens the door for many future investigations. We leave these issues open for future studies.

\section{Introductory Examples}
\label{sectionGeneral}

To set the stage, recall that we are interested in extremal correlators of chiral primary operators (CPOs) in four-dimensional SCFTs on $\mathbb{R}^4$ in perturbation theory (that is, in the sector with zero instanton number). Moreover, we will concentrate on Lagrangian ${\cal N}=2$ theories with freely generated chiral rings based on a simple gauge group. The latter contains a scalar $\phi$.
It is convenient to introduce the notation: 
\begin{equation}
\label{defPhik}
    \phi_k \equiv {\rm Tr}\phi^k\ .
\end{equation}
Note that the range of possible $k$'s is set by the gauge algebra, for instance, in $\mathrm{SU}(N)$, $k=2,\cdots,N$ (the operators $\phi_j$ with $j>N$ can be expressed in terms of the fundamental invariants).

Due to superconformal invariance, the computation of correlation functions of CPOs on $\mathbb{R}^4$ can be mapped to the computation of correlation functions in the $\mathbb{S}^4$ matrix model. However one must disentangle the additional mixing of operators when going to the $\mathbb{S}^4$ due to the conformal anomaly. As argued in \cite{Gerchkovitz:2016gxx}, this can be done by a Gram-Schmidt orthogonalization. Let us first study this procedure in the simplest $\mathrm{SU}(2)$ and $\mathrm{SU}(3)$ cases,  which  already 
exhibit some key features.

\subsection{The $\mathrm{SU}(2)$ case}

Let us consider, for definiteness, superconformal QCD (SQCD) with gauge group $\mathrm{SU}(2)$
(that is, $\mathrm{SU}(2)$ theory with 4 fundamental massless hypermultiplets). Note that results for $\mathrm{SU}(2)$ $\mathcal{N}=4$ SYM can be easily extracted by simply considering the leading term in the perturbative
expansion in powers of $g$ (this is because ${\cal N}=4$ correlators are not renormalized).

For $\mathrm{SU}(2)$ we have a single Casimir invariant $\phi_2$. Hence one would expect that the CPOs form a one-dimensional tower $O_n=\phi_2^n$ starting with the identity. However, when going to the $\mathbb{S}^4$, due to the conformal anomaly, $O_n$, $O_m$ for $n\ne m$ are not anymore orthogonal. The mixture can be disentangled by a Gram-Schmidt procedure, which in this case amounts to writing 

\begin{equation}
    O_0=1\,,\qquad O_1=\phi_2-\alpha_1^0\,O_0\,,\qquad O_2=\phi_2^2-\alpha_2^1\,O_1-\alpha_2^0\,O_0\,,\qquad \cdots
\end{equation}
where the $\alpha_i^j$ are easily fixed by demanding orthogonality of the $O_i$'s with all the other $O_j$ for $j<i$ with respect to the inner product defined by the $\mathbb{S}^4$ matrix model. Then, by construction $\langle O_n|O_m\rangle = \widetilde{G}_{n}\, \delta_{nm}$, so that the desired correlators on $\mathbb{R}^4$ are obtained by simply normalizing $G_n=\frac{\widetilde{G}_n}{\widetilde{G}_0}$. As a consequence of this structure, and as first anticipated in \cite{Baggio:2014sna}, these correlators satisfy a Toda equation of the form

\begin{equation}
    \partial_{\tau} \partial_{\bar{\tau}} \log \widetilde{G}_n=\frac{\widetilde{G}_{n+1}}{\widetilde{G}_n}-\frac{\widetilde{G}_{n}}{\widetilde{G}_{n-1}}\,.
\end{equation}
This can be put in a more familiar form by a defining $G_n =e^{q_n}$. Then one has a system of equations
representing a  Toda chain \cite{Baggio:2014sna}.
The emergence of this Toda chain can be deduced from the $tt^*$ equations of \cite{Papadodimas:2009eu}. An alternative approach, relying on localization, was presented in \cite{Gerchkovitz:2016gxx}. As discussed in that reference, the Gram-Schmidt procedure can be efficiently encoded as a ratio of certain determinants. In turn, since the (perturbative) matrix model only depends on $\tau,\ \bar\tau $ (the YM coupling) through the classical contribution, which is $Z_{cl}=e^{-2\pi\,{\rm Im}\tau\,\phi_2}$,  derivatives with respect to $\tau,\ \bar\tau $ give rise to further insertions of $\phi_2$ in those determinants. As shown in \cite{Gerchkovitz:2016gxx}, this leads to the emergence of the Toda equation described above (we refer to \cite{Gerchkovitz:2016gxx} for the complete argument).

Note that the argument leading to the Toda equation solely relies on the algebraic structure of the Gram-Schmidt orthogonalization plus the fact that, discarding instantons, the only dependence on $\tau$ is through $Z_{cl}$, so that $\tau$, $\bar\tau$-derivatives of the partition function give rise to insertions of $\phi_2$. Thus the Toda equation holds for any SCFT with gauge group $\mathrm{SU}(2)$.

\subsection{The $\mathrm{SU}(3)$ case}

Let us now consider the case of SQCD with $\mathrm{SU}(3)$ gauge group, again restricting to perturbation theory. Just as in the $\mathrm{SU}(2)$ case, the $\mathcal{N}=4$ correlators can be easily extracted as the leading terms in the $g\rightarrow 0$ limit. 

Results for the first few operators in this case can be found in \cite{Baggio:2015vxa,Gerchkovitz:2016gxx}. In this case the Casimir invariants are $\phi_2$ and $\phi_3$. Hence the operators resulting from the diagonalization process starting with a monomial $\phi_2^n\phi_3^m$ can be labelled as $O_{(n,m)}$. The $\mathrm{SU}(3)$ case exhibits a new feature with respect to the $\mathrm{SU}(2)$ case.
Since the procedure described in \cite{Gerchkovitz:2016gxx} is designed to remove mixtures with lower-dimensional operators
arising from the anomalous $\mathbb{S}^4\leftrightarrow \mathbb{R}^4$ conformal mapping, when there is degeneracy (\textit{i.e.} more than one operator at a given dimension), the matrix of correlators is typically non-diagonal. Indeed, for $\mathrm{SU}(3)$ SQCD this happens starting at dimension 6, where one has the operators $\phi_2^3$ and $\phi_3^2$. Let us first focus on the correlators for the lowest dimensional operators where this issue does not arise, in particular on $\widetilde{G}_{(0,0)}$, $\widetilde{G}_{(1,0)}$ and $\widetilde{G}_{(2,0)}$. It is straightforward to check that these satisfy the Toda equation\footnote{There is a factor of $4^{\Delta}$ with respect to \cite{Gerchkovitz:2016gxx} reabsorbed in the normalization of the operators as in \cite{Baggio:2015vxa}.}
\begin{equation}
    \partial_{\tau}\partial_{\bar{\tau}}\log\widetilde{G}_{(1,0)}=\frac{\widetilde{G}_{(2,0)}}{\widetilde{G}_{(1,0)}}-\frac{\widetilde{G}_{(1,0)}}{\widetilde{G}_{(0,0)}}\, .
\end{equation}
In fact, it is easy to compute these correlators to very high loop orders and check that the Toda equation holds (we have done this up to tenth loop order $g^{20})$. Motivated by this, it is reasonable to wonder whether the Toda equation should hold generically. However, one immediately encounters the problem alluded above, namely, that  correlators are not diagonal since the diagonalization has only been performed with respect to lower dimensional operators. In order to remedy this, we could add one further step on top of the procedure of \cite{Gerchkovitz:2016gxx} and iteratively construct the operators by orthogonalizing with respect to all others with lower dimension, and as well those of the same dimension.\footnote{This strategy was adopted in \cite{Beccaria:2018xxl}  for the operators
 $({\rm Tr}\phi^2)^n{\rm Tr}\phi^3$ in $\mathrm{SU}(3)$ and $({\rm Tr}\phi^2)^n{\rm Tr}\phi^4$ in $\mathrm{SU}(4)$.} Note that, as in any orthogonalization procedure, this bears a certain ambiguity, since one has to choose an ordering to run the Gram-Schmidt algorithm. In order to devise the optimal strategy, and with an eye on the Toda equation, note that given an operator it gives rise to a full tower upon insertions of $\phi_2$, which in turn is slightly special since its insertions, at the perturbative level, can be traded by derivatives with respect to $\tau$ (which is, together with the algebra of the Gram-Schmidt procedure, why the Toda equation emerged for $\mathrm{SU}(2)$). This suggests to momentarily step back and consider instead organizing the operators as in table \eqref{tabOperatorsSU3}.

\begin{figure}
    \centering
\begin{tabular}{cccccccccccccc}
  Operator/$\Delta$ & 0 & 1 & 2 & 3 & 4 & 5 & 6 & 7 & 8 & 9 & 10 & \dots \\ \hline 
  $(n,0)$  & 1 &  & $\phi_2$ &  & $\phi_2^2$ &  & $\phi_2^3$ &  & $\phi_2^4$ &  & $\phi_2^5$ &   \dots\\ 
  $(n,1)$  &  &  &  & $\phi_3$ &  & $\phi_2\phi_3 $ &  & $\phi_2^2\phi_3 $ &  & $\phi_2^3\phi_3 $ &    & \dots\\
  $(n,2)$  &  &  &  & & &   & $\phi_3^2 $ &  & $\phi_2\phi_3^2 $ &  & $\phi_2^2\phi_3^2 $  & \dots \\
  $(n,3)$  &  &  &  &  &  &  & &  &  & $\phi_3^3$ &    & \dots \\   &  &  &  &  &  & $\vdots$  &$\vdots$  &  &  &  &  &    &  
\end{tabular}
\caption{Ordering the operators in $\mathrm{SU}(3)$ SQCD }
    \label{tabOperatorsSU3}
\end{figure}

The ordering in table \eqref{tabOperatorsSU3} suggests an strategy for such orthogonalization: choose a $\Delta_{\textrm{max}}$ and orthogonalize the operators  starting from those on the first row, from left to right, until the dimension is higher than $\Delta_{\textrm{max}}$. Then move to the next row and iterate, including previous towers, until exhausting all operators of dimension smaller or equal than the chosen $\Delta_{\textrm{max}}$. 

Below we quote the first few correlators (up to $\Delta_{\textrm{max}}=6$) computed in this way (recall that, by construction, the matrix of correlators is diagonal). We introduce 
$F_{(m,n)}$ defined by
\begin{equation}
\label{definitionF}
  G_{(m,n)} = \left(\frac{g^2}{16\pi}\right)^{\Delta_{(m,n)}}\, F_{(m,n)}\ .
  \end{equation}

\noindent {\it $\mathrm{SU}(3)$ superconformal theory with 6 fundamental hypermultiplets:}
\begin{eqnarray}
{F}_{(0,0)}&=&1\,, \nonumber \\ \nonumber \\
{F}_{(1,0)}&=& 16-\frac{45 \zeta (3) g^4}{2 \pi ^4}+\frac{425 \zeta (5) g^6}{8 \pi ^6}+\frac{25 \left(1188 \zeta (3)^2-3577 \zeta (7)\right) g^8}{768 \pi ^8}+\mathcal{O}(g^{10})\, ,\nonumber \\ \nonumber \\
{F}_{(2,0)}&=&640-\frac{2160 \zeta (3) g^4}{\pi ^4}+\frac{6375 \zeta (5) g^6}{\pi ^6}+\frac{25 \left(24516 \zeta (3)^2-67963 \zeta (7)\right) g^8}{96 \pi ^8}+\mathcal{O}(g^{10})\, ,\nonumber \\ \nonumber  \\
{F}_{(3,0)}&=&46080-\frac{272160 \zeta (3) g^4}{\pi ^4}+\frac{969000 \zeta (5) g^6}{\pi ^6}
+\frac{15 \left(325296 \zeta (3)^2-876365 \zeta (7)\right) g^8}{4 \pi ^8}+\mathcal{O}(g^{10})\, ,\nonumber \\ \nonumber \\
{F}_{(0,1)}&=&40-\frac{135 g^4 \zeta (3)}{2 \pi ^4}+\frac{6275 g^6 \zeta (5)}{48 \pi^6}+\frac{25 g^8 \left(7452 \zeta (3)^2-15533 \zeta (7)\right)}{1536 \pi^8} +\mathcal{O}(g^{10}) \, , \nonumber \\
{F}_{(1,1)}&=&1120-\frac{4410 g^4 \zeta (3)}{\pi ^4}+\frac{144725 g^6 \zeta (5)}{12 \pi^6}+\frac{35 g^8 \left(157356 \zeta (3)^2-350665 \zeta (7)\right)}{384 \pi^8}+\mathcal{O}(g^{10})\, , \nonumber \\
{F}_{(0,2)}&=&6720-\frac{28350 \zeta (3) g^4}{\pi ^4}+\frac{139125 \zeta (5) g^6}{2 \pi ^6}+\frac{1575 \left(477 \zeta (3)^2-854 \zeta (7)\right) g^8}{8 \pi ^8}+\mathcal{O}(g^{10})\, . \nonumber 
\end{eqnarray}
It is straightforward to compute these correlators to an arbitrary order. Note that already at this 4-loop order there is non-linear dependence on the  Riemann $\zeta$ function coefficients. We also note that  $F_{(1,1)}$  correctly reproduces the correlator previously computed in (3.16) in \cite{Beccaria:2018xxl}.  

One can now check that the Toda equations\footnote{Recall that the $\tilde G_{(n,m)}$'s represent the unnormalized correlators, with $G_{(n,m)}=\tilde G_{(n,m)}/\tilde G_{(0,0)}$. } 
\begin{equation}
\label{TodaSU3}
    \partial_{\tau}\partial_{\bar{\tau}}\log\widetilde{G}_{(n,m)}=\frac{\widetilde{G}_{(n+1,m)}}{\widetilde{G}_{(n,m)}}-\frac{\widetilde{G}_{(n,m)}}{\widetilde{G}_{(n-1,m)}}\, ,
\end{equation}
are satisfied. We have checked this system of equations up to, and including 12-loop order $O(g^{24})$. 

Note that, just as in the $\mathrm{SU}(2)$ case, the emergence of the Toda equations only relies on the algebra of the orthogonalization process plus the fact that insertions of $\phi_2$ can be traded by derivatives with respect to $\tau$. Hence the Toda equation above should hold in any SCFT with gauge group $\mathrm{SU}(3)$. Following the classification of \cite{Koh:1983ir}, reviewed in section \ref{classif}, there are three such theories, namely the SQCD case which we have just studied, the $\mathcal{N}=4$ case (where the result holds trivially, since it is akin to keeping only the leading terms in $g$ in the SQCD case) and $\mathrm{SU}(3)$ with one rank-2 symmetric and one fundamental hypermultiplet (see next subsection). For this latter case, proceeding as described, one finds the correlators
\begin{eqnarray}
{F}_{(0,0)}&=&1\, ,\nonumber \\ \nonumber \\ 
{F}_{(1,0)}&=&16-\frac{25 \zeta (5) g^6}{4 \pi ^6}+\frac{6125 \zeta (7) g^8}{192 \pi ^8}-\frac{33075 \zeta (9) g^{10}}{256 \pi ^{10}}  \nonumber \\ && +\frac{175 \left(740 \zeta (5)^2+35497 \zeta (11)\right) g^{12}}{12288 \pi ^{12}}+\mathcal{O}(g^{14})\,,\nonumber  \\ \nonumber  \\ 
{F}_{(2,0)}&=&640-\frac{750 \zeta (5) g^6}{\pi ^6}+\frac{116375 \zeta (7) g^8}{24 \pi ^8}-\frac{99225 \zeta (9) g^{10}}{4 \pi ^{10}}+ \nonumber \\ && \nonumber \\ && +\frac{125 \left(5160 \zeta (5)^2+248479 \zeta (11)\right) g^{12}}{256 \pi ^{12}}+\mathcal{O}(g^{14})\,,\nonumber \\ \nonumber \\
{F}_{(3,0)}&=&46080-\frac{114000 \zeta (5) g^6}{\pi ^6}+\frac{900375 \zeta (7) g^8}{\pi ^8}-\frac{22524075 \zeta (9) g^{10}}{4 \pi ^{10}}+ \nonumber \\ \nonumber \\ && +\frac{375 \left(116740 \zeta (5)^2+5715017 \zeta (11)\right) g^{12}}{64 \pi ^{12}}+\mathcal{O}(g^{14})\,,\nonumber \\ \nonumber \\ 
{F}_{(0,1)}&=&40-\frac{925 g^6 \zeta (5)}{24 \pi ^6}+\frac{67375 g^8 \zeta (7)}{384 \pi^8}-\frac{341775 g^{10} \zeta (9)}{512 \pi ^{10}} \nonumber \\ \nonumber \\ && +g^{12} \left(\frac{649375
   \zeta (5)^2}{9216 \pi ^{12}}+\frac{23321375 \zeta (11)}{9216 \pi^{12}}\right)+O\left(g^{14}\right)\,,\nonumber \\ \nonumber \\ 
{F}_{(1,1)}&=&1120-\frac{19075 g^6 \zeta (5)}{6 \pi ^6}+\frac{1941625 g^8 \zeta (7)}{96 \pi^8}-\frac{13307175 g^{10} \zeta (9)}{128 \pi ^{10}} \nonumber \\ \nonumber \\ && +g^{12}  \left(\frac{30265375 \zeta (5)^2}{2304 \pi ^{12}}+\frac{2378674375 \zeta
   (11)}{4608 \pi ^{12}}\right)+O\left(g^{14}\right)\,,\nonumber \\ \nonumber \\ 
{F}_{(0,2)}&=&6720-\frac{18375 \zeta (5) g^6}{\pi ^6}+\frac{1708875 \zeta (7) g^8}{16 \pi ^8}-\frac{32645025 \zeta (9) g^{10}}{64 \pi ^{10}}+ \nonumber \\ \nonumber \\ && +\frac{6125 \left(2405 \zeta (5)^2+100089 \zeta (11)\right) g^{12}}{256 \pi ^{12}}+\mathcal{O}(g^{14})\, . \nonumber \\
\label{SU(3)+S+F}
\end{eqnarray}
Again,  here we explicitly quote results up to the first non-linear order in the $\zeta$'s. It is nevertheless straightforward to go to any desired higher order, checking that indeed, the Toda equations are satisfied for the corresponding $\tilde G_{(n,m)}$.

Remarkably, \eqref{SU(3)+S+F} does not contain terms with $\zeta(3)$. This arises due to a surprising cancellation in the two-loop contribution to the partition function of the coefficient of $\zeta(3)$. Hence, any supersymmetric observable in this theory that can be computed from insertions in the localized (matrix model) partition function will not have, in perturbation theory, any contribution proportional to any power of $\zeta(3)$.

\subsection{A lesson}\label{lesson}

It is instructive to come back to the ordering prescription to run the Gram-Schmidt orthogonalization.  
For instance, up to dimension $\Delta_{\textrm{max}} = 8$, in the $\mathrm{SU}(3)$ case such sequence is
\begin{equation}
    (0,0)\rightarrow (1,0) \rightarrow (2,0) \rightarrow (3,0) \rightarrow (4,0) \rightarrow(0,2) \rightarrow (1,2)\, .
\end{equation}
To begin with, one may wonder what would happen if one chose a different ordering. Consider, for example, 
\begin{equation}
    (0,0)\rightarrow (1,0) \rightarrow (2,0) \rightarrow (3,0) \rightarrow (0,2) \rightarrow(4,0) \rightarrow (1,2)\, .
\end{equation}
In this case, one would find that the decoupled Toda equations fail to hold. Here operators have been arranged in order of increasing dimension, but this is not the correct order that leads to decoupled Toda chains. Explicit calculations will be given in section \ref{sectionComparisonPrecision}. 

It should also be noted that our sequence of orthogonalization implies that a given operator $O_I \phi_2^n$ may mix with a higher dimensional
component of the upper tower.\footnote{We thank Bruno Le Floch for useful comments on that point.}
When this happens, identifying the flat-theory counterpart of the resulting orthogonal operators is less straightforward, particularly in considering the cutoff $\Delta_{\textrm{max}} \to\infty $. Nonetheless, it is important to note that the corresponding correlation functions --which are the relevant physical observables-- are well defined and satisfy Toda equation (1.2) (mixing coefficients, instead, suffer from holomorphic ambiguities \cite{Gerchkovitz:2016gxx}). 

Interestingly, the universal Toda structure can already be exhibited with no need of a full orthogonalization of the CPOs
belonging to different towers.
The main point is that a Toda chain exists in each tower of operators of the form  $O_I \phi_2^n$.
Once these operators are orthogonalized through GS procedure by arranging them in order of increasing dimension,
the resulting correlation functions satisfy the same Toda chain equation (1.2) independently of the seed $O_I$. In this way, one can exhibit the Toda structure with no need of mixing with higher dimensional operators. Of course, the complete orthogonal basis is eventually needed in order to determine the complete set of  $\mathbb{R}^4$ correlation functions of CPOs.\footnote{In general, different sequences in GS orthogonalization may give different correlation functions.
This seems to reflect the ambiguity in the normal ordering prescription in defining $\mathbb{R}^4$ composite operators; see e.g. \cite{Billo:2017glv} for calculations in the SQCD context.}

A proof is as follows. Consider  the first tower $\{ \phi_2^n\} $.
In \cite{Bourget:2018obm}, in the context of $SU(N)$ SQCD, $N\leq 5$,  it was shown that orthogonalizing  these operators by arranging them in order of increasing dimension leads to the same Toda equation (\ref{TodaGeneral}). The underlying reason can be understood from the form  of the partition function in the zero-instanton sector,
\begin{equation}
Z = \int [da]  e^{-2\pi {\rm Im}\tau \phi_2} 
f\left( \phi_2,\phi_3,\phi_4,...\right)\ .
\end{equation}
Two-point correlation functions of $\phi_2^n$ are obtained by differentiation with respect to $\tau $. As a result,
the final diagonal correlators are given by the determinant formula
(see \cite{Gerchkovitz:2016gxx}), which is known to satisfy (\ref{TodaGeneral}).
Next, consider a tower with seed $O_I$, i.e. $\{ O_I \phi_2^n\} $.
Suppose we orthogonalize only operators belonging to this tower. So we consider correlators
with a single insertion of $O_I O_I$ and  insertions $\phi_2^n\ \phi_2^m$.
Since our proof for the first tower $\{ \phi_2^n\} $  does not rely on the specific form of $f$, 
clearly the same proof applies now, replacing $f$ by 
$g(\phi_2,\phi_3,...)\equiv O_I O_I f(\phi_2,\phi_3,...) $.

Having shown that each orthogonalized tower satisfies Toda equations (\ref{TodaGeneral}), the next problem is to complete the orthogonalization among operators belonging to  different towers {\it without spoiling the Toda structure of the correlators}. This is highly non-trivial, because correlators will change once each operator gets mixed with operators of other towers. We claim that the Toda structure is maintained by our ordering described above (see general definition in section 3.1). On the contrary, a sequence of orthogonalization where one orthogonalizes a given operator with respect to {\it all} operators of lower dimensions (including those belonging to different towers) fails already at three loops, as shown in  \cite{Gerchkovitz:2016gxx}. Orthogonalizing with respect to operators of {\it lower or equal} dimensions, as in \cite{Beccaria:2018xxl}, also fails (see section \ref{sectionComparisonPrecision}). This justifies our choice of ordering given above.

In general, there may be more than one tower at a given dimension.
For the cases with $\mathrm{SU}(2)$ and $\mathrm{SU}(3)$ gauge groups discussed above this 
degeneration does not occur.
A simple example is given by the towers corresponding respectively to  $\phi_3^4$ and to $\phi_4^3$ in $\mathrm{SU}(N)$ for $N\geq 4$. As  described below, we find that, when these degenerate towers occur, either ordering between them leads to the same decoupled Toda chains structure.
Our claim is that the (non-normalized) correlators $\widetilde{G}_I=\langle O_I\,O_I\rangle$ obtained using this ordering of operators satisfy the infinite set of Toda equations (\ref{TodaSU3}) and that these can be packaged in the compact form (\ref{TodaGeneral}). 

It is also important to stress that the algorithm is an \textit{orthogonalization} and not an \textit{orthonormalization}. The normalization of the operators is already fixed by the convenient choice of (coupling independent) three-point functions, which take values 1 or 0.
 
Note that the above prescription is independent of the gauge group. In the next section, we show that the algorithm generalizes for arbitrary Lagrangian SCFTs based on a simple gauge group.

\section{Extremal Correlators and Generalized Toda Equation}
\label{algorithm}

\subsection{The Toda Orders and the Main Equation}

Let us consider any four-dimensional (Lagrangian) $\mathcal{N}=2$ CFT with a simple gauge group (therefore, we consider a connected Lie group). We refer to appendix \ref{AppendixNotations} for notations. 
The chiral ring is freely generated, which means that we can choose a (linear) basis of monomial operators that we call $(\phi_{I})$ labeled by a multi-index $I \in \mathcal{I}:=\mathbb{N}^r$, defined by 
\begin{equation}
    \phi_I = \prod\limits_{k \in D(\mathfrak{g})} \phi_k^{I_k} \, . 
\end{equation}
This last equation crucially uses the fact that the ring is freely generated: any monomial operator admits a \emph{unique} expression in terms of the $\phi_k$, so the family of operators $(\phi_I)_{I \in \mathcal{I}}$ is indeed a linear basis. The chiral ring structure constants are defined by the OPE 
\begin{equation}
    \phi_I (x) \phi_J (0) = \sum\limits_{K \in \mathcal{I}} C_{IJ}^K \phi_K (0) + \dots  \, . 
\end{equation}
Thanks to the monoid structure of $\mathcal{I}$, the structure constants are trivial, 
\begin{equation}
    C_{IJ}^K = \delta_{I+J,K} \, . 
\end{equation}
We define a family of total orders, that we call the \emph{Toda orders}, on the set of indices $\mathcal{I}$ as follows. Let us denote by $\mathcal{S}$ the map that associates to each $I = (n_1 , \dots , n_r) \in \mathcal{I}$ with its seed, 
\begin{equation}
    \mathcal{S}(I) = (0,n_2,\dots,n_r) \, . 
\end{equation}
Then the order is defined by $I<I'$ if 
\begin{itemize}
    \item $\Delta (\mathcal{S}(I))<\Delta (\mathcal{S}(I'))$ or
    \item $\Delta (\mathcal{S}(I))=\Delta (\mathcal{S}(I'))$ and $\mathcal{S}(I) \prec \mathcal{S}(I')$ or
    \item $\mathcal{S}(I) = \mathcal{S}(I')$ and $n_1<n_1'$. 
\end{itemize}
In this definition, $\prec$ is any total order on the $(r-1)$-tuples, and this is why there are several Toda orders when the rank of the gauge group is $\geq 3$. The smallest non-vanishing element is $(1,0,\dots ,0)$ which we denote by the shorthand $\mathbf{1}$. These Toda orders formalize what we described in the previous section: operators are organized in towers of the form 
\begin{equation}
    T_I=\{O_I,\,O_I\,\phi_2,\,O_I\,\phi_2^2,\,\cdots\}\, , 
\end{equation}
and the towers are ordered by the conformal dimension of their seed. When various seeds have the same dimension, we order them arbitrarily using $\prec$ (one can choose e.g. the lexicographic order). 

We can deform the theory on the sphere $\mathbb{S}^4$, introducing parameters $\tau_k$ for $k \in D(\mathfrak{g})$. Here $D(\mathfrak{g}) \in \mathbb{N}^r$ is the set of degrees of fundamental invariants of $\mathfrak{g}$, and we identify $\tau \equiv \tau_2$. The partition function is (compare with the undeformed partition function in Appendix \ref{AppendixNotations}) 
\begin{equation}
  Z^{\mathcal{T}} [\tau_k , \bar{\tau}_k] = \int_{\mathfrak{h}} [\mathrm{d} a] \Delta (a)  Z_{\mathrm{1-loop}} ^{\mathcal{T}} (a) \exp \left(- \sum\limits_{k \in D(\mathfrak{g})} 2 \pi^{k/2}   {\rm Im} \, \tau_k \,   \phi_k \right)   \, . 
\end{equation}
Likewise, we introduce the differential operators 
\begin{equation}
    \partial_I = \prod\limits_{k \in D(\mathfrak{g})} \left( \frac{\partial}{\partial \tau_{k}} \right)^{I_k} \, ,  \qquad  \bar{\partial}_I = \prod\limits_{k \in D(\mathfrak{g})} \left( \frac{\partial}{\partial \bar{\tau}_{k}} \right)^{I_k} \, .   
\end{equation}
Then one can compute the (infinite) matrix of \emph{unnormalized} correlators of operators on $\mathbb{S}^4$, which is given by derivatives of the sphere partition function $Z^{\mathcal{T}} [\tau_k , \bar{\tau}_k]$ with respect to the couplings and setting $\tau_{k'} = \bar{\tau}_{k'} = 0$ for $k' \in D(\mathfrak{g})-\{2\}$. We call this matrix $\widetilde{M}$, so that 
\begin{equation}
    \widetilde{M}_{IJ} = \left( \partial_I \bar{\partial}_J Z^{\mathcal{T}}  \right) |_{\tau_{k'>2} = \bar{\tau}_{k'>2} = 0}  \,.
\end{equation}
We then apply an unnormalized Gram-Schmidt orthogonalization process to $\widetilde{M}$, in the order defined above on $\mathcal{I}$. The eigenvalues of the matrix obtained in that way are denoted $\widetilde{G}_I  = \widetilde{G}_I (\tau , \bar{\tau})$. We claim that these obey the generalized Toda equation 
\begin{equation}\tag{1.2}
\partial_{\tau}\partial_{\bar{\tau}}\log\widetilde{G}_I=\frac{\widetilde{G}_{I+\mathbf{1}}}{\widetilde{G}_{I}}-\frac{\widetilde{G}_{I}}{\widetilde{G}_{I-\mathbf{1}}}\,. 
\end{equation}
Note that this equation is valid for all $I \in \mathcal{I}$ such that $I-\mathbf{1}$ exists (i.e. which is not a seed $\mathcal{S}(I')$).\footnote{The Toda equation presented here can be seen as a generalization of the usual semi-infinite Toda chain, the difference lying in the structure of the ordered set of indices $\mathcal{I}$. While the usual semi-infinite Toda chain is labeled by the integers $\mathbb{N}$, which as a totally ordered set form the infinite ordinal $\omega$, equation (\ref{TodaGeneral}) is defined on $\mathcal{I}$, which is the ordinal $\omega^2$. } Therefore, it must be supplemented with boundary conditions, and extremal correlators of CPOs depend on the theory under study only through these boundary conditions and the degrees of fundamental invariants.

\subsection{Finite-dimensional implementation}

In practice, for concrete computations in perturbation theory, one has to implement orthogonalization up to some maximal conformal dimension $\Delta_{\textrm{max}}$. Because of the structure of the Toda orders, $\Delta_{\textrm{max}}$ has to be chosen carefully depending on the order of expansion in perturbation theory: if the sphere partition function is computed to precision $O(g^{2d})$, then one needs 
\begin{equation}
    \Delta_{\textrm{max}} \geq d \, . 
\end{equation}
Then the algorithm can be formalized as follows:
\begin{enumerate}
    \item List all the operators $\phi_I$ with conformal dimension $\Delta \leq \Delta_{\textrm{max}}$. 
    \item Order these operators following a Toda order. 
    \item Compute the matrix $\tilde{M}$ in that basis. 
    \item Perform a Gram-Schmidt orthogonalization algorithm (without normalization) on $\widetilde{M}$. 
\end{enumerate}
The diagonal elements of the matrix thus obtained are the correlators $\widetilde{G}_I$.

\begin{figure}
    \centering \scriptsize
\begin{tabular}{ c ccccccccccc c c}
  Operator /$\Delta$&  0 & 1 & 2 & 3 & 4 & 5 & 6 & 7 & 8 & 9 & 10 & 11 & 12  \\ \hline 
  $(n,0,0)$  & 1 &  & $\phi_2$ &  & $\phi_2^2$ &  & $\phi_2^3$ &  & $\phi_2^4$ &  & $\phi_2^5$ &  & $\phi_2^6$ \\ 
  $(n,1,0)$  &  &  &  & $\phi_3$ &  & $\phi_2\phi_3 $ &  & $\phi_2^2\phi_3 $ &  & $\phi_2^3\phi_3 $ & & $\phi_2^4\phi_3 $ &  \\
  $(n,0,1)$  &  &  &  &  & $\phi_4$ &  & $\phi_2\phi_4$ &  & $\phi_2^2\phi_4$ &  & $\phi_2^3\phi_4$ & &$\phi_2^4\phi_4$ \\
  $(n,2,0)$  &  &  &  & & &   & $\phi_3^2 $ &  & $\phi_2\phi_3^2 $ &  & $\phi_2^2\phi_3^2 $ & & $\phi_2^3\phi_3^2 $ \\
  $(n,1,1)$  &  &  &  &  &  &  & & $\phi_3\phi_4$ &  & $\phi_2\phi_3\phi_4$ &  & $\phi_2^2\phi_3\phi_4$ &  \\ 
  $(n,0,2)$  &  &  &  &  &  &  & &  & $\phi_4^2$ & &  $\phi_2 \phi_4^2$ & & $\phi_2^2 \phi_4^2$ \\
   $(n,3,0)$  &  &  &  &  &  &  & &  &  & $\phi_3^3$ &  & $\phi_3^3\phi_2$ &  \\  
   $(n,2,1)$  &  &  &  &  &  &  & &   &  &  & $\phi_3^2\phi_4$ & & $\phi_3^3\phi_4$  \\ 
     $(n,4,0)$  &  &  &  &  &  &  & &   &  &  &  & & $\phi_3^4$  \\ 
      $(n,0,3)$  &  &  &  &  &  &  & &   &  &  &  & & $\phi_4^3$  \\ 
\end{tabular}
    \caption{Schematic representation of the operators $\phi_I$ for $\mathrm{SU}(4)$ theory (rank 3). The ordering of the operators is obtained by going through the (infinite) lines one after another. In practical computations, one introduces a cut-off $\Delta_{\textrm{max}}$ on the right; then the number of operators is finite, and the orthogonalization should be done line after line (and not column after column). While immaterial to obtain the Toda structure, we have chosen an ordering for $\phi_3^4$ and $\phi_4^3$. }
    \label{tabOperatorsSU4}
\end{figure}

\subsection{Comparison with other orderings}
\label{sectionComparisonPrecision}

We stress that the algorithm   presented here is not only sufficient to obtain the Toda equation, but it is also necessary, in the sense that any ordering that is not a Toda order will fail to give extremal correlators obeying the decoupled Toda Chains (\ref{TodaGeneral}).
In the cases where no two seeds have the same conformal dimension, there is a unique Toda order and we claim
that this unique Toda order is the only ordering that gives correlators satisfying  (\ref{TodaGeneral}).\footnote{On the other hand, we have checked on one example the degree of freedom left by the choice of $\prec$. Namely, we computed the first terms of the two chains at dimension $\Delta=12$ in the $\mathrm{SU}(4)$ theory (see Figure \ref{tabOperatorsSU4}) of type \textbf{(A5)} -- defined in section \ref{classif} -- and checked, in a 16-loop computation, that the two orderings of $(0,4,0)$ and $(0,0,3)$ are both compatible with (\ref{TodaGeneral}). }

To conclude this section, we will now 
carry out
 some precision tests that allow to compare the correlators obtained with different orderings.
In order to perform very high loop order calculations, one trick is to formally replace some of the $\zeta(2n-1)$ in equation (\ref{zetass}) by $0$; as a by-product, the results have a more manageable size and can be reported here to high enough precision so that the effects of the orderings appear. Here we consider $SU(3)$ SQCD theory 
and  focus on the terms in the perturbation series involving only  $\zeta(5)$ coefficients, which can be achieved by formally replacing $\zeta(2n-1)$ by $0$ for any $n \neq 3$. 
We analyze in this theory the influence of the order for implementing the GS
procedure on the correlators labeled $(4,0)$, $(0,2)$, $(1,2)$, $(2,2)$. We  compare 
\begin{enumerate} 
    \item The (unique) Toda order 
        \begin{equation}
        \{  (n,0) | n \in \mathbb{N} \} \cup \{ (0,2) , (1,2) , (2,2)\}  \, . 
    \end{equation}
    The correlators computed using this order will be denoted with the letter $F$, following the convention (\ref{definitionF}). 
    \item Arranging operators in order of increasing conformal dimension
        \begin{equation}
        \{ (0,0),(1,0),(2,0),(3,0),(0,2),(4,0) , (1,2) , (5,0), (2,2) \} \,.  
    \end{equation}
    The correlators computed using this order will be denoted $F'$. 
\end{enumerate} 
We find 
    \begin{eqnarray}
       F_{(4,0)} &=& 5160960+\frac{194208000 g^6 \zeta (5)}{\pi
   ^6}+\frac{6686216250 g^{12} \zeta (5)^2}{\pi
   ^{12}}+O\left(g^{15}\right) \nonumber \\ \nonumber
      F'_{(4,0)} &=& 5160960+\frac{194208000 g^6 \zeta (5)}{\pi
   ^6}+\frac{6685625625 g^{12} \zeta (5)^2}{\pi
   ^{12}}+O\left(g^{15}\right) \\ \nonumber 
    F_{(0,2)} &=& 6720+\frac{139125 g^6 \zeta (5)}{2 \pi ^6}+\frac{39265625 g^{12} \zeta (5)^2}{48 \pi^{12}}+O\left(g^{18}\right) \\ \nonumber 
           F'_{(0,2)} &=& 6720+\frac{139125 g^6 \zeta (5)}{2 \pi ^6}+\frac{838534375 g^{12} \zeta (5)^2}{1024 \pi ^{12}}+O\left(g^{18}\right) \\
    F_{(1,2)} &=& 268800+\frac{6478500 g^6 \zeta (5)}{\pi ^6}+\frac{900878125 g^{12} \zeta (5)^2}{6 \pi
   ^{12}}+ O\left(g^{18}\right) \\ \nonumber
    F'_{(1,2)} &=&  268800+\frac{6478500 g^6 \zeta (5)}{\pi ^6}+\frac{9623009375 g^{12} \zeta (5)^2}{64
   \pi ^{12}}+ O\left(g^{18}\right) \\ \nonumber
    F_{(2,2)} &=& 23654400+\frac{1043196000 g^6 \zeta (5)}{\pi ^6}+\frac{119937702500 g^{12} \zeta (5)^2}{3 \pi
   ^{12}}+ O\left(g^{18}\right) \\ \nonumber 
    F'_{(2,2)} &=& 23654400+\frac{1043196000 g^6 \zeta (5)}{\pi ^6}+\frac{640773739375 g^{12} \zeta 
   (5)^2}{16 \pi ^{12}}+ O\left(g^{18}\right) 
    \end{eqnarray}
We see that the correlators obtained by the two orderings differ
in the 6 loop term.
Using these, we can compute the degree of violation of the decoupled Toda equations for both orders: 
    \begin{eqnarray}
                 \left.    \partial_{\tau}\partial_{\bar{\tau}}\log\widetilde{G}_I- \left( \frac{\widetilde{G}_{I+\mathbf{1}}}{\widetilde{G}_{I}}-\frac{\widetilde{G}_{I}}{\widetilde{G}_{I-\mathbf{1}}} \right)\right\lvert_{I=(3,0)}  &=&O\left(g^{18}\right)  \nonumber  \\ \nonumber 
       \left.     \partial_{\tau}\partial_{\bar{\tau}}\log\widetilde{G}'_I- \left( \frac{\widetilde{G}'_{I+\mathbf{1}}}{\widetilde{G}'_{I}}-\frac{\widetilde{G}'_{I}}{\widetilde{G}'_{I-\mathbf{1}}} \right)\right\lvert_{I=(3,0)}  &=&  \frac{13125 g^{16} \zeta (5)^2}{262144 \pi
   ^{14}}+O\left(g^{18}\right)  \\ 
          \left.            \partial_{\tau}\partial_{\bar{\tau}}\log\widetilde{G}_I- \left( \frac{\widetilde{G}_{I+\mathbf{1}}}{\widetilde{G}_{I}}-\frac{\widetilde{G}_{I}}{\widetilde{G}_{I-\mathbf{1}}} \right)\right\lvert_{I=(1,2)}  &=&O\left(g^{18}\right)  \\ 
          \left.         \partial_{\tau}\partial_{\bar{\tau}}\log\widetilde{G}'_I- \left( \frac{\widetilde{G}'_{I+\mathbf{1}}}{\widetilde{G}'_{I}}-\frac{\widetilde{G}'_{I}}{\widetilde{G}'_{I-\mathbf{1}}} \right)\right\lvert_{I=(1,2)}  &=& -\frac{28125 g^{16} \zeta (5)^2}{262144 \pi ^{14}}+ O\left(g^{18}\right)  \nonumber
    \end{eqnarray}
Therefore, we find that (\ref{TodaGeneral}) is satisfied for a Toda order only.

\section{Further Examples}

\subsection{General ${\cal N}=2$ CFTs with gauge group $\mathrm{SU}(N)$}\label{classif}

As discussed in \cite{Koh:1983ir}, for theories based on an $\mathrm{SU}(N)$ gauge group there are essentially 8 cases which we list below, explicitly constructing the 1-loop factor for each case. We first recall that the contribution to the 1-loop partition function of the (adjoint) vector multiplet is

\begin{equation}
    Z_{\textrm{1-loop}}^{\rm (VM)}=\prod_{i<j}H(a_i-a_j)^2\ ,
\end{equation}
where $H(x)$ is defined in (\ref{defH}).
In addition we will need the contributions of hypermultiplets in various representations: the fundamental (of dimension $N$ and Dynkin label $[1,\,0\,,\cdots,\,0]$), the rank-2 symmetric representation (of dimension $N(N+1)/2$ and Dynkin label $[2,\,0,\,\cdots,\,0]$), the rank-2 antisymmetric representation (of dimension $N(N-1)/2$ and Dynkin label $[0,\,1,\,\cdots,\,0]$) and the rank-3 antisymmetric representation (of dimension $\frac{1}{3!}N(N-1)N-2)$ and  Dynkin label $[0,\,0,\,1,\,\cdots,\,0]$). These read, respectively,

\begin{eqnarray}
&&Z_{\textrm{1-loop}}^{\rm (fund)}=\prod_{i}\frac{1}{H(a_i)}\,, \nonumber \\ \nonumber  \\ &&
Z_{\textrm{1-loop}}^{\rm (2-symm)}=\frac{1}{\prod_i H(2a_i) \prod_{i<j} H(a_i+a_j) }\,, \nonumber \\ \nonumber \\ &&    Z_{\textrm{1-loop}}^{\rm (2-antisymm)}=\frac{1}{\prod_{i<j} H(a_i+a_j) }\,, \nonumber \\ \nonumber \\ &&      Z_{\textrm{1-loop}}^{\rm (3-antisymm)}=\frac{1}{ \prod_{i<j<k} H(a_i+a_j+a_k) }\, .
\end{eqnarray}

\noindent{\bf (A1)} $\mathcal{N}=4$ SYM. This case is familiar enough and we will refer to the literature for the explicit formulas.

\medskip

\noindent{\bf (A2)} $2N$ fundamental representations (SQCD) \cite{Howe:1983wj}. This case has been extensively studied and we will refer to the literature for the explicit formulas.

\noindent{\bf (A3)} $N-2$ fundamental representations and one rank-2 symmetric representation. This case occurs for all $N\geq 3$.
The complete one-loop factor is given by 
\begin{equation}
    Z_{\textrm{1-loop}} =\frac{\prod_{i<j}H(a_i-a_j)^2}{\left[\prod_i H(a_i)^{N-2} H(2a_i)\right] \prod_{i<j} H(a_i+a_j) }\ .
\end{equation}
One can check that the exponential factor $e^{-x^2/n}$ in the function $H(x)$, defined in (\ref{defH}),
cancels out, as it should be for a conformal field theory. This factor will cancel, as expected, in all examples below. In theories where this factor does not cancel out (e.g. ${\cal N}=2$ $\mathrm{SU}(N)$ theory with $N_f<2N$ fundamentals), there is a logarithmic UV divergence in the original one-loop determinants that has to be absorbed into a renormalization of the coupling constant, leading to a non-zero $\beta $ function \cite{Pestun:2007rz}.

\medskip

\noindent{\bf (A4)} 
\smallskip

\noindent a) $N+2$ fundamental representations and one rank-2 antisymmetric representation. This case occurs for all $N\geq 4$. The complete one-loop factor in the localization partition function is given by

\begin{equation}
    Z_{\textrm{1-loop}} = \frac{\prod_{i<j}H(a_i-a_j)^2}{\left(\prod_i H(a_i)\right)^{N+2} \prod_{i<j} H(a_i+a_j) }\ .
\end{equation}
A discussion on large $N$ properties of this theory can be found in \cite{Fiol:2015mrp}.
\smallskip

\noindent b)  CFT with two rank-2 antisymmetric and 4 fundamental representations, occurring for all $N\geq 5$.

\begin{equation}
    Z_{\textrm{1-loop}} = \frac{\prod_{i<j}H(a_i-a_j)^2}{\left(\prod_i H(a_i)\right)^{4} \prod_{i<j} H(a_i+a_j)^2 }\ .
\end{equation}

\noindent{\bf (A5)} The CFT with one rank-2 symmetric and one rank-2 antisymmetric, for $N\geq 4$. This gives the one-loop factor

\begin{equation}
    Z_{\textrm{1-loop}} = \frac{\prod_{i<j}H(a_i-a_j)^2}{\left(\prod_i H(2a_i)\right) \prod_{i<j} H(a_i+a_j)^2 }\ .
\end{equation}

Remarkably, for these theories, the two-loop term, proportional to $\zeta(3)$, exactly cancels out. Hence we find a ``non-renormalization" theorem for all $\mathrm{SU}(N)$ CFTs with one rank-2 symmetric and one rank-2 antisymmetric representations: the two-loop contribution to any extremal correlator of CPOs
vanishes.
This can be seen by using equation (\ref{zetass}) of the appendix and noting that the $\zeta(3)$ term
cancels between numerator and denominator. 
It also has implications for any
supersymmetric observable that can be computed from the localized partition function, including, in particular, 
the VEV of the 1/2 BPS circular Wilson loop and correlation functions between the Wilson loop operator and CPOs
\cite{Rodriguez-Gomez:2016cem,Billo:2018oog}: the corresponding perturbation series do not contain $\zeta(3)$ coefficients
and the two loop terms are the same as in ${\cal N}=4$ super Yang-Mills.
In the perturbation theory computed with
ordinary methods, this requires a massive cancellation of Feynman diagrams.
This is a surprise, since the theory should not have any additional supersymmetry.
We do not understand the underlying reason for this two-loop cancellation. 
Presumably, it could be due to the fact  that the group-theoretic factors in some combined Feynman diagrams
accidentally coincide with the case of the hypermultiplet in the adjoint  representation
(note that the matter content of this theory is similar
to that of the ${\cal N}$=4 $\mathrm{SU}(N)\times \mathrm{U}(1)$ theory, since both theories have the same number of hypermultiplets, $\frac{1}{2}N(N+1)+\frac{1}{2}N(N+1)=N^2$). Clearly, it would be interesting to understand the origin of this cancellation.

Note that the case (A3) with $N=3$ is also included in this family, since for $N=3$ the antisymmetric representation is equivalent to the fundamental representation.

\medskip

\noindent{\bf (A6)}
The CFT with two rank-3 antisymmetric,
occurring only for $\mathrm{SU}(6)$. We obtain
\begin{equation}
    Z_{\textrm{1-loop}}^{\mathrm{SU}(6)} = \frac{\prod_{i<j}H(a_i-a_j)^2}{\prod_{i<j<k} H(a_i+a_j+a_k)^2}\ .
\end{equation}

\medskip

\noindent{\bf (A7)} A CFT with one rank-3 antisymmetric and   $N_f=\frac12 (9N-N^2-6)$, appearing  for $N=6,7,8$ (for lower $N$, it becomes equivalent to one of the above CFTs).
The one-loop factor is

\begin{equation}
    Z_{\textrm{1-loop}}^{\mathrm{SU}(N)} = \frac{\prod_{i<j}H(a_i-a_j)^2}{\left(\prod_i H(a_i)\right)^{N_f} \prod_{i<j<k} H(a_i+a_j+a_k)}\ ,\qquad (N_f,N)=(6,6),(4,7),(1,8)\ .
\end{equation}

\medskip

\noindent{\bf (A8)}
An $\mathrm{SU}(6)$ gauge theory with two fundamental representations, one rank-2 antisymmetric and one rank-3 antisymmetric. We have
\begin{equation}
    Z_{\textrm{1-loop}}^{\mathrm{SU}(6)} = \frac{\prod_{i<j}H(a_i-a_j)^2}{\left(\prod_i H(a_i)\right)^{2} \prod_{i<j} H(a_i+a_j)\prod_{i<j<k} H(a_i+a_j+a_k)}\ .
\end{equation}
Once again, we check that the exponential factor $e^{-x^2/n}$ in the $H(x)$ functions cancel in all cases, as expected.
\medskip

\subsubsection*{Toda equation}

We have checked that, within the limits allowed by the computational capabilities, upon implementing our algorithm the  correlation functions  of all the $\mathrm{SU}(N)$ theories listed above  satisfy the universal Toda equations \eqref{TodaGeneral}.
Our checks include up to $N=5$ and beyond ten-loop order.\footnote{
We omit the long formulas for the correlators in each case, which are kindly available upon request.} 

In the following subsections, we provide some explicit examples for other cases: symplectic and orthogonal gauge groups.

\subsection{Symplectic gauge group}

As outlined above, we expect our procedure to hold independently of the theory, and in particular, for any gauge group. 

As an example, we will consider $\mathrm{USp}(4)$ SQCD, that is
${\cal N}=2$ SYM with gauge group $\mathrm{USp}(4)$ and  6 hypermultiplets in the fundamental representation. 
Using the general formula \eqref{zzzone} in appendix, we find that the 1-loop determinant for ${\cal N}=2$ super Yang-Mills with gauge group $\mathrm{USp}(2N)$  and $2N+2$ fundamental hypermultiplets is given by
\begin{equation}
    Z_{\textrm{1-loop}} = \frac{\prod_{1 \leq i \leq N} H(2 a_i)^2 \prod_{1 \leq i < j \leq N}  H(a_i+a_j)^2 H(a_i-a_j)^2}{\prod_{1 \leq i \leq N} H(a_i)^{2(2N+2)}}\ .
\end{equation}
The group $\mathrm{USp}(4)$, of type $C_2$, has a chiral ring generated by operators with degrees 2 and 4, so that our operators will be labelled as $O_{(n,m)}$.  Using the orthogonalization algorithm, we can compute the correlation functions. For illustrative purposes, here we show  the first three non-trivial terms of the Toda chain with seed $(0,0)$ (here and below, we use again functions $F$ defined analogously to (\ref{definitionF})):
\begin{eqnarray}
{F}_{(1,0)} &=& 20-\frac{945 g^4 \zeta (3)}{32 \pi ^4}+\frac{17325 g^6 \zeta (5)}{256 \pi ^6}+\frac{36225 g^8 \left(48   \zeta (3)^2-133 \zeta (7)\right)}{32768 \pi ^8}+\mathcal{O}(g^{10})\, ,\nonumber \\ \nonumber \\
{F}_{(2,0)} &=& 960-\frac{6615 g^4 \zeta (3)}{2 \pi ^4}+\frac{294525 g^6 \zeta (5)}{32 \pi ^6}+\frac{6615 g^8 \left(6168
   \zeta (3)^2-15295 \zeta (7)\right)}{4096 \pi ^8}+\mathcal{O}(g^{10})\, ,\nonumber \\ \nonumber \\
  {F}_{(3,0)} &=& 80640-\frac{476280 g^4 \zeta (3)}{\pi ^4}+\frac{12525975 g^6 \zeta (5)}{8 \pi ^6} + \nonumber \\ \nonumber \\ && +\frac{19845 g^8
   \left(110136 \zeta (3)^2-260015 \zeta (7)\right)}{1024 \pi ^8}+\mathcal{O}(g^{10})\,, \nonumber \\ \nonumber 
\end{eqnarray}
and the first three terms of the Toda chain with seed $(0,1)$: 
\begin{eqnarray}
  {F}_{(0,1)} &=& 105-\frac{33075 g^4 \zeta (3)}{128 \pi ^4}+\frac{1126125 g^6 \zeta (5)}{2048 \pi ^6}  \nonumber \\ \nonumber \\ && +\frac{33075 g^8
   \left(4614 \zeta (3)^2-9443 \zeta (7)\right)}{262144 \pi ^8}+\mathcal{O}(g^{10})\,,\nonumber \\ \nonumber\\
{F}_{(1,1)} &=& 3780-\frac{297675 g^4 \zeta (3)}{16 \pi ^4}+\frac{27286875 g^6 \zeta (5)}{512 \pi ^6}+ \nonumber \\ \nonumber \\ && +\frac{99225 g^8
   \left(46602 \zeta (3)^2-101479 \zeta (7)\right)}{65536 \pi ^8}+\mathcal{O}(g^{10})\,.\nonumber \\ 
   {F}_{(2,1)} &=& 302400-\frac{2381400 g^4 \zeta (3)}{\pi ^4}+\frac{66268125 g^6 \zeta (5)}{8 \pi ^6} \nonumber \\    \nonumber
   & &   +\frac{496125 g^8   \left(54423 \zeta (3)^2-120232 \zeta (7)\right)}{2048 \pi ^8}   +O\left(g^{10}\right)\,.\nonumber 
\end{eqnarray}
Once again, one can check that the corresponding unnormalized correlation functions 
$\tilde G_{(n,m)}$ satisfy the universal Toda equation \eqref{TodaGeneral}.

\medskip

On the other hand, implementing orthogonalization
by arranging operators in order of increasing conformal dimension
would again lead  to a failure of the Toda equation at four loops.
In particular, one would find $ F_{(3,0)}-F'_{(3,0)} = \frac{8505 g^8 \zeta (3)^2}{16 \pi
   ^8}+...$, and \eqref{TodaGeneral} would be violated by terms of order $\zeta(3)^2g^{12}$ for $I=(2,0)$.

\subsection{Orthogonal gauge group}

In this final subsection,  we give one example for an ${\cal N}=2$ SCFT with an orthogonal gauge group of type $B_N$. From \eqref{zzzone}, we find that
the one-loop determinant for SQCD (where there are $2N-1$ hypermultiplets in the fundamental representation) is 
\begin{equation}
    Z_{\textrm{1-loop}} = \frac{\prod_{1 \leq i \leq N} H(a_i)^2 \prod_{1 \leq i < j \leq N}  H(a_i+a_j)^2 H(a_i-a_j)^2}{\prod_{1 \leq i \leq N} H(a_i)^{2(2N-1)}}
\end{equation}
We choose $SQCD$ with gauge group $\mathrm{SO}(7)$ -- a rank 3 case.  We only quote the first few orders for the first two Toda chains. Again, one can check that \eqref{TodaGeneral} is satisfied in all cases. 

\begin{eqnarray} 
{F}_{(0,0,0)} &=& 1 \, ,\nonumber \\
{F}_{(1,0,0)} &=& 42-\frac{945 g^4 \zeta (3)}{32 \pi ^4}+\frac{17325 g^6 \zeta (5)}{256 \pi
   ^6}+\frac{4725 g^8 \left(192 \zeta (3)^2-917 \zeta (7)\right)}{32768 \pi
   ^8}+O\left(g^{10}\right) \, ,\nonumber \\
{F}_{(2,0,0)} &=& 3864-\frac{23625 g^4 \zeta (3)}{4 \pi ^4}+\frac{121275 g^6 \zeta (5)}{8 \pi
   ^6} \nonumber \\ && +\frac{4725 g^8 \left(921 \zeta (3)^2-3668 \zeta (7)\right)}{512 \pi
   ^8}+O\left(g^{10}\right) \, ,\nonumber \\
{F}_{(3,0,0)} &=& 579600-\frac{5740875 g^4 \zeta (3)}{4 \pi ^4}+\frac{130717125 g^6 \zeta (5)}{32
   \pi ^6}    \\ && +\frac{42525 g^8 \left(6383856 \zeta (3)^2-22799371 \zeta
   (7)\right)}{94208 \pi ^8}+O\left(g^{10}\right) \, ,\nonumber
\end{eqnarray}

\begin{eqnarray}
{F}_{(0,1,0)} &=& \frac{15120}{23}-\frac{1068795 g^4 \zeta (3)}{2116 \pi ^4}+\frac{17307675 g^6
   \zeta (5)}{16928 \pi ^6}+O\left(g^{8}\right) \, ,\nonumber \\
{F}_{(1,1,0)} &=& \frac{876960}{23}-\frac{32180085 g^4 \zeta (3)}{529 \pi ^4}+\frac{646932825 g^6
   \zeta (5)}{4232 \pi ^6}+O\left(g^8\right)  \, , \\
{F}_{(2,1,0)} &=& \frac{108743040}{23}-\frac{6325888590 g^4 \zeta (3)}{529 \pi ^4}+\frac{72579501225
   g^6 \zeta (5)}{2116 \pi ^6}+O\left(g^8\right) \, . \nonumber
\end{eqnarray}

\section*{Acknowledgements}

We would like to thank Matteo Beccaria and Bruno Le Floch for useful correspondence. A.B. and  D.R-G are supported by the Spanish Government grant MINECO-16-FPA2015-63667-P. J.G.R.  acknowledges financial support from projects 2017-SGR-929, MINECO grant FPA2016-76005-C2-1-P.

\begin{appendix}

\section{Notations and conventions}
\label{AppendixNotations}


We consider a Lagrangian theory $\mathcal{T}$ with a simple\footnote{This implies in particular that the group is connected. } gauge group with Lie algebra $\mathfrak{g}$ of rank $r$ and a matter content that makes it a CFT. 
The sphere partition function of the theory $\mathcal{T}$ on $\mathbb{S}^4$ is given by the localization formula \cite{Pestun:2007rz} 
\begin{equation}
\label{localizationFormulaWithInstantons}
  Z^{\mathcal{T}}_{S^4} [\tau , \bar{\tau}] = \int_{\mathfrak{h}} [\mathrm{d} a] \Delta (a) Z_{\mathrm{1-loop}} ^{\mathcal{T}}(a)  \exp \left( - 2 \pi  {\rm Im} \, \tau \,   \phi_2 \right) Z_{\mathrm{inst}}. 
\end{equation}
where $\phi_2$ is the generator of the chiral ring at degree 2 (see the normalization (\ref{defPhik})), $\mathfrak{h}$ is the Cartan subalgebra of $\mathfrak{g}$, $\Delta (a)$ is the Vandermonde determinant 
\begin{equation}
     \Delta (a)  = \prod\limits_{\beta \in \mathrm{Roots}^+(\mathfrak{g})} (\beta \cdot a)^2 \, ,  
\end{equation}
and $Z_{\mathrm{1-loop}} ^{\mathcal{T}}(a)$ is the one-loop determinant defined by 
\begin{equation}
\label{zzzone}
     Z_{\mathrm{1-loop}} ^{\mathcal{T}}(a) = \frac{\prod\limits_{\beta \in \mathrm{Roots}(\mathfrak{g})}H( \beta \cdot a)}{\prod\limits_{w\in \mathrm{Weights}(\mathcal{T})} H(w \cdot a)}
\end{equation}
with 
\begin{equation}
\label{defH}
H(x)\equiv \prod_{n=1}^{\infty}\Big(1+\frac{x^2}{n^2}\Big)^{n} e^{-\frac{x^2}{n}}\, .
\end{equation}
In this paper we study the sector with zero instanton number, so  we  set $Z_{\mathrm{inst}} = 1$. Perturbation theory is generated by using the expansion
\begin{equation}
\label{zetass}
    \log H(x)= -\sum_{n=2}^\infty (-1)^n \frac{\zeta(2n-1)}{n}\, x^{2n}\ .
\end{equation}

\end{appendix}

\end{document}